\pdfoutput=1
\documentclass[conference]{IEEEtran}
\usepackage{cite}
\ifCLASSINFOpdf
  \usepackage[pdftex]{graphicx}
  \graphicspath{{pdf/}{img/}}
  \DeclareGraphicsExtensions{.pdf,.png,.jpg.eps}
\else
  \usepackage[dvips]{graphicx}
  \graphicspath{{../eps/}}
  \DeclareGraphicsExtensions{.eps}
\fi
\usepackage[cmex10]{amsmath}
\usepackage{balance}
\usepackage{array}
\usepackage[caption=false,font=footnotesize]{subfig}
\usepackage{url}

\usepackage{eurosym}
\usepackage{tabularx}
\usepackage{epstopdf}
\newtheorem{thm}{Theorem}
\newtheorem{cor}[thm]{Corollary}

\def \paperTitle {Analysis of Downlink and Uplink Decoupling in Dense Cellular Networks}
\def \journalTitle {}

\def \submissionVolume {X}
\def \submissionNumber {Y}
\def \publicationDate {MONTH~YEAR}
\begin{document}
\title{\paperTitle}

\author{\IEEEauthorblockN{Alexis~I.~Aravanis, Olga Mu{\~n}oz, Antonio Pascual-Iserte and Josep Vidal}
\IEEEauthorblockA{Dept. of Signal Theory and Communications,\\
Universitat Polit{\`e}cnica de Catalunya (UPC),\\
Barcelona, Spain\\
Email: \{alexios.aravanis, olga.munoz, antonio.pascual, josep.vidal\}@upc.edu}
}

\markboth{\journalTitle,~Vol.~\submissionVolume, No.~\submissionNumber, \publicationDate}%
{\firstAuthorSurname \MakeLowercase{\textit{et al.}}: \PaperTitle}
\maketitle

\begin{abstract}
Decoupling uplink (UL) and downlink (DL) is a new architectural paradigm where DL and UL are not constrained to be associated to the same base station (BS). Building upon this paradigm, the goal of the present paper is to provide lower, albeit tight bounds for the ergodic UL capacity of a decoupled cellular network. The analysis is performed for a scenario consisting of a macro BS and a set of small cells (SCs) whose positions are selected randomly according to a Poisson point process of a given spatial density. Based on this analysis simple bounds in closed form expressions are defined. The devised bounds are employed to compare the performance of the decoupled case
versus a set of benchmark cases, namely the coupled case, and the situations of having either a single macro BS or only SCs. This comparison provides valuable insights regarding the behavior and performance of such networks, providing simpler expressions for the ergodic UL capacity as a function of the distances to the macro BS and the density of SCs. These expressions constitute a simple guide to the minimum degree of densification that guarantees the Quality of Service (QoS) objectives of the network, thus, providing a valuable tool to the network operator of significant practical and commercial value.
\end{abstract}


\IEEEpeerreviewmaketitle

\section{Introduction}
\label{sec:introduction}
\IEEEPARstart{T}he ever increasing demand for broadband access, together with the current shift from asymmetric traffic loads to symmetric traffic applications (i.e. symmetric as to the downlink (DL) and uplink (UL) traffic), has induced a change in the archetypal perception of heterogeneous networks (HetNets). In particular, the advent of applications such as social media or on-line video gaming applications resulted in an unabated increase in the UL traffic, mandating a dedicated optimization of the UL channel. 

In this course, a new architectural paradigm emerged allowing for the standalone management of UL and DL connectivity. Downlink and uplink decoupling (DUDe) \cite{DUD1, DUD2,DUD3,DUD4} allows the optimal network operation by addressing UL and DL as separate network connections. Thus, each user equipment (UE) can be connected to a different serving node in the UL and the DL. The feasibility of this approach relies on the density of the current HetNets and the disparity between the transmit power of the UE, the small cells (SCs) and the legacy macro cells (MCs).  

Specifically, the differences in the transmit power of the SCs, the MCs and the UE could result in a setting where the connection to a MC -transmitting at a high power- would maximize the DL user rate, whereas the maximization of the UL user rate would impose the connection to a SC -residing in the vicinity of the user- given the restricted power of the UE. Thus, the independent management of the UL connectivity in DUDe arises as an additional pillar of flexibility, providing substantial capacity \cite{DUD4,TWC} and power \cite{Power} gains. Moreover, the DUDe is a more general approach accounting also for the entrenched coupled association policy, since the latter can be viewed as a restrictive special case of the more general DUDe scheme \cite{ACK}.

The flexibility provided by the DUDe is fully in line with the framework of the \textquotedblleft device-centric\textquotedblright~architecture where the connectivity is provided dynamically by network nodes based on the particularities of each specific device during each specific session \cite{DUD3}. Thus, the limelight is gradually shifted from the MCs of the single tier-homogeneous networks, through the SCs of the multi-tier heterogeneous networks to the device end of the envisaged 5G networks. Moreover, this shift is following the general trend of network densification and cell-less philosophy. In particular, the deployment of ultra-dense cellular networks provides substantial capacity gains, leading to a remarkable increase of the average UE throughput \cite{Ultra-dense}.  

Building upon the benefits of DUDe and of network densification, the contribution of the present paper is twofold. At the outset, focusing on DUDe we employ lower, albeit tight bounds, for the ergodic UL capacity to provide closed form expressions for quantifying the performance of the network. The obtained analytical expressions, facilitate the comparison of the network performance under a DUDe association policy against the performance of the cases corresponding to a network comprising a single MC, a multitude of SCs, or the classical DL-UL coupled philosophy. This comparison provides valuable insights into the system behavior, which in turn allows for the devise of simpler analytical models to characterize the expected UL rate. 

Subsequently, as opposed to relevant attempts hitherto focusing solely on the characterization of the UL channel \cite{TWC,Reviewer}, the present paper exploits the devised simple analytical models to provide insight into the minimum degree of densification of SCs, that guarantees meeting the Quality of Service (QoS) objectives. Specifically, even though it is known that there exists an upper limit for the densification of a 5G ultra-dense network, imposed by the backhaul capacity and energy constraints \cite{Backhaul}, the current work focuses on the lower limit of the densification of SCs guaranteeing that the QoS objectives will be achieved. Thus, this sets out a densification road map for the network operator and designer of significant practical and commercial value.

The remainder of the paper is organized as follows. Section \ref{sec:Ergodic} presents the considered network architecture comprising a MC and SCs. Furthermore, analytical expressions are defined calculating the lower, yet tight bounds for the UL ergodic capacity in the case of the network comprising only a MC, in the case of the network comprising only SCs and in the general case comprising both under DUDe and coupled association policies. The obtained closed form expressions associate the average user capacity with the density of SCs in the network. Section \ref{sec:simulations} presents the simulation results corroborating that the devised bounds from the preceding analysis are tight. Finally, Section \ref{sec:conclusion} concludes the paper and presents perspectives.

\section{The Wireless Cellular Network Architecture and the UL Ergodic Capacity Bounds}
\label{sec:Ergodic}

\subsection{The Wireless Cellular Network Scenario}
\begin{figure}[h]
  \begin{center}
   \centerline{
    \includegraphics[height=3cm]{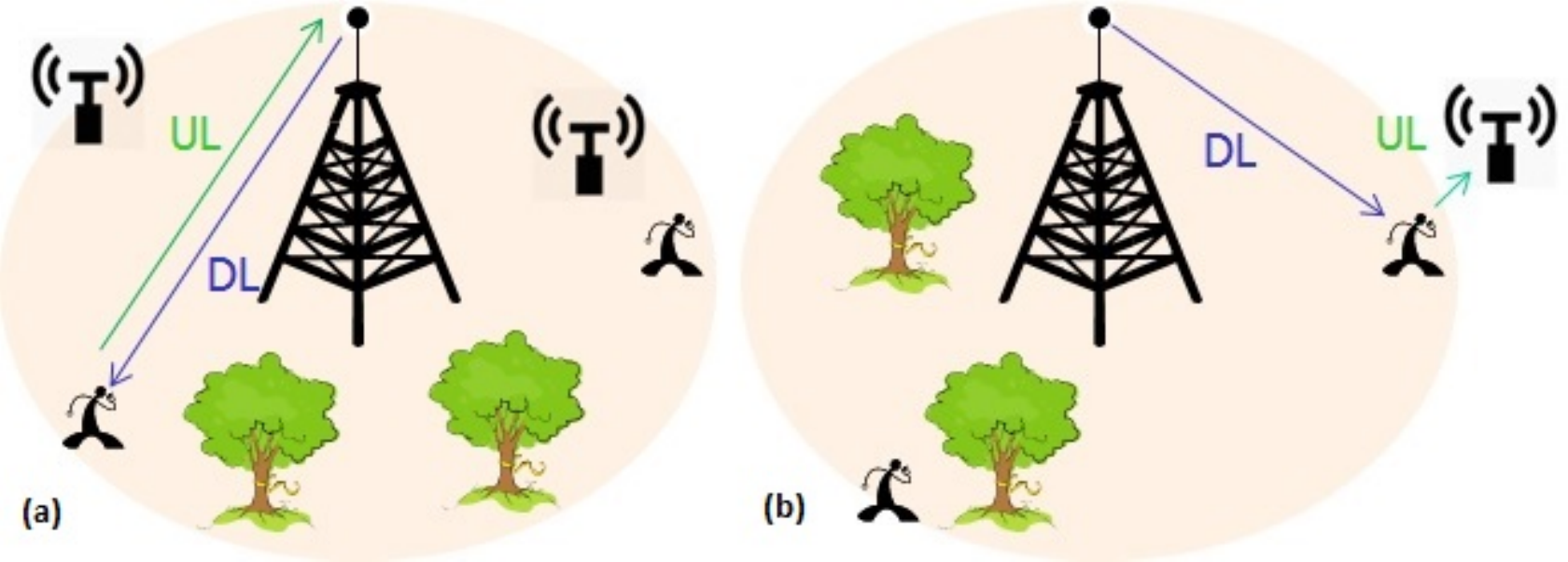}}
    \caption{Indicative scenario under a coupled and a DUDe association policy.}
    \label{fig:Pic1}
  \end{center}
\end{figure}

A wireless cellular system is considered, comprising a MC served by the access point $AP_0$. Moreover, a set of SCs are overlaid by means of low power and low complexity access points $AP_i$, whose positions are generated according to a homogeneous Poisson point process (PPP) of density $\lambda$ $(SCs / m^2)$ \cite{PPP}. $AP_0$ transmits at a high power level. On the other hand, all $AP_i$ transmit at a low power level. Furthermore, for the sake of simplicity in the notation, it is assumed that all UEs and access points, both for the MC and the SCs, are equipped with one antenna. However, the extension to the multi-antenna case is straightforward and in this course, a relevant analysis is provided throughout the paper when necessary. In particular, this analysis elaborates on how the expressions should be changed when considering $M$ antennas in the macro access point and $N$ antennas in the SCs access points. An illustration of the above scenario is presented for a coupled and a DUDe association policy in Fig. (\ref{fig:Pic1}a) and Fig. (\ref{fig:Pic1}b) respectively.
Intra cell users are assumed to be sharing orthogonal resources, as is the typical case in literature \cite{TWC}, whereas adjacent SCs are assumed to coordinate, using different operating frequencies and thus, providing an interference free scenario. The feasibility of this inter-cell interference free approach relies on two pillars. On the one hand, on the low transmit power level of UEs in the UL and, on the other hand, on the interference mitigating effect of blocking in mmWave bands \cite{Blocking}. The latter ensures that the extent of the necessary coordination to provide an interference free scenario is finite and thus, feasible.

This intrinsic counteraction of the interference motivates the extension of the analysis to account for blocking and remains to be tackled in future work. To elaborate, the SCs coordination assumption and the blocking effect can give rise to a novel approach where interference free meta-cells (i.e. clusters) of SCs emerge based on blocking. The borders of the non-interfering meta-cells are defined by the blocking effect allowing for a frequency reuse among meta-cells and thus, providing an interference free setting for ultra dense networks. In other words, it can be assumed that SCs within each meta-cell do not interfere due to coordination, and SCs outside the meta-cell do not interfere due to blocking.

\subsection{UL Ergodic Capacity Bounds - Simplified Cases}
This section focuses on the derivation of a lower bound on the ergodic UL capacity for different setups, including the DUDe scheme. In this course, it is taken into account that, in general, the selection of the serving $APs$ does not follow the changes of the fast fading. Therefore, the selection of the $AP$ is averaged over fast fading.

\subsubsection{Ergodic Capacity for a Single MC Network}
\label{sssec:MC}
The ergodic capacity of a user placed at distance $d_0$ from $AP_0$ resulting from the average over fast fading, if no additional SC access points $AP_i$ are overlaid, is given by
\begin{eqnarray}
\label{eqn:MC}
\mathbf{E}[R] = \mathbf{E}_{h_0}[\log(1 + d_0^{-\beta} {h_0}^2 \gamma)],
\end{eqnarray}
where the expectation is with respect to the fading coefficient $h_0$, assuming a Rayleigh fading where $h_0$ follows a zero-mean circularly symmetric Gaussian distribution with variance equal to $1$. $\log(·)$ in all the expressions henceforth represents the natural logarithm, $\beta$ is the path-loss exponent \cite{Goldsmith}, and $\gamma$ is the $\textrm{SNR}$ at the reference distance with
\begin{eqnarray}
\label{eqn:gamma}
\gamma = \frac{P_{UE}}{\sigma^2 L_{ref}}.
\end{eqnarray}
In (\ref{eqn:gamma}) $P_{UE}$ is the transmission power of the UE, $\sigma^2$ is the noise power, and $L_{ref}$ is the equivalent path-loss at a reference distance of $1$ meter, which includes also the effects of the transmit and receive antenna gains. 

A lower, albeit very tight bound for the ergodic capacity can be derived from \cite{Eduard} as follows:
\begin{eqnarray}
\label{eqn:Bound}
\mathbf{E}_{h_0}[\log(1 + d_0^{-\beta} {h_0}^2 \gamma)] = ~~~~~~~~~~~~~~  \nonumber \\
\mathbf{E}_{h_0}[\log(1 + d_0^{-\beta} \gamma \exp(\log({h_0}^2)))] \geq ~  \nonumber \\
\log(1+  d_0^{-\beta} \gamma \exp(\mathbf{E}_{h_0}[\log({h_0}^2)]))=~ \nonumber \\
\log(1 + d_0^{-\beta}\gamma~\rho). ~~~~~~~~~~~~~~~~~~~~~~~~~
\end{eqnarray}
where the inequality in (\ref{eqn:Bound}) arises from Jensen's inequality and the convexity of the $\log(1+\exp(x))$ function \cite{Boyd}. For Rayleigh fading, $\rho$ is the expectation of the logarithm of a Chi-square random variable which is equal to \cite{35}: 
\begin{eqnarray}
\label{eqn:Rayleigh}
\rho = \exp(\mathbf{E}_{h_0}[\log{h_0}^2]) = \exp(-\psi),
\end{eqnarray}
where $\psi \simeq 0.577$ is the Euler-Mascheroni constant \cite{36}.

In the case of a multi-antenna receiver, the preceding analysis holds with $h_0$ being replaced by $\lVert{\mathbf {h_0}}\rVert^2$. $\mathbf {h_0}$ is a vector composed by $n$ i.i.d. elements, each one corresponding to the Rayleigh fading coefficient between the transmitter and the $n\textsuperscript{th}$ antenna receiver with a variance equal to $1$. Moreover, (\ref{eqn:Rayleigh}) needs to be revised accordingly \cite{35}, with $\rho$ in the case of a multi-antenna receiver being equal to: 
\begin{eqnarray}
\label{eqn:RayleighGeneral}
\rho(n) = \exp\left(\mathbf{E}_{\mathbf {h_0}}[\log\lVert{\mathbf {h_0}}\rVert^2]\right) = \exp\left(-\psi + \sum_{j=1}^{n-1} \frac{1}{j}\right).
\end{eqnarray}
Hence, in case the access point at the MC is equipped with $M$ antennas, the above expression is calculated for $\rho(M)$, and in case the access points at the SCs employ $N$ antennas, for $\rho(N)$. However, for the present analysis assuming single-antenna access points and according to (\ref{eqn:Rayleigh}) $\rho$ is employed, whereas brief guidelines are provided throughout the paper toward adapting the expressions to the multi-antenna case whenever needed.

\subsubsection{Ergodic Capacity for a Network of Small Cells}
\label{sssec:SC}
The ergodic capacity of a user residing within a dense deployment of SCs served only by the access points $AP_i$ can be considered to be independent of the position of the user within the coverage of the network. That is, due to the assumption that the access points $AP_i$ are deployed according to a homogeneous PPP with spatial density $\lambda$. Assuming a distance $d$ between a reference user and the $AP$ of the closest SC, the probability density function (PDF) of the distance $d$ between a reference user and its closest $AP$ is given by \cite{PDF}
\begin{eqnarray}
\label{eqn:CDF}
f_d(d) = 2 \pi d \lambda \exp(-\lambda \pi d^2),
\end{eqnarray}
where $d \geq 0$.

Thus, we can again calculate a lower bound of the ergodic capacity for the case of a network consisting only of SCs. In this case, it is expected that the UE will connect to the closest SC and the expectation is with respect to both the fading and the distance $d$ to the closest $AP$. Thus, proceeding as in (\ref{eqn:Bound}) a bound of the ergodic capacity is given by
\begin{eqnarray}
\label{eqn:SC}
\mathbf{E}_{h,d}[\log(1 + d^{-\beta} {h}^2 \gamma)] = ~~~~~~~~~~~~~~~~~~~~~~~  \nonumber \\
\mathbf{E}_{h,d}[\log(1 + \exp(\log (d^{-\beta} {h}^2 \gamma)))] \geq ~~~~~~~~~~~  \nonumber \\
\log(1 + \gamma \exp(\mathbf{E}_{h,d}[\log(d^{-\beta}{h}^2 )])) = ~~~~~~~~~~  \nonumber \\
\log(1+ \gamma \exp(- \beta \mathbf{E}_{d}[\log(d)] + \mathbf{E}_h[\log({h}^2)])).
\end{eqnarray}
where, according to (\ref{eqn:Rayleigh}), $\exp(\mathbf{E}_h[\log({h}^2)]) = \rho$ and, according to (\ref{eqn:CDF}), the expected value $\mathbf{E}_{d}[\log(d)]$ can be computed as follows:
\begin{eqnarray}
\label{eqn:PDF}
\mathbf{E}_{d}[\log(d)] = \int_{0}^{\infty} \log(r) 2 \pi r \lambda \exp(-\lambda \pi r^2) dr = ~~ \nonumber \\
2 \int_{0}^{\infty} x \log(x) \exp(-x^2)dx - \log(\sqrt{\pi \lambda}) = ~~~~~~~ \nonumber \\
-\frac{\psi}{2} - \log(\sqrt{\pi \lambda}). ~~~~~~~~~~~~~~~~~~~~~~~~~~~~~~~~~~~~~~~
\end{eqnarray}
Thus, combining (\ref{eqn:SC}) and (\ref{eqn:PDF}) a bound for the ergodic capacity is obtained as follows:
\begin{eqnarray}
\label{eqn:SC1}
\mathbf{E}_{h,d}[\log(1 + d^{-\beta} {h}^2 \gamma)] \geq \log\left(1 + \gamma (\lambda \pi)^{\frac{\beta}{2}} \rho \exp( \beta \frac{\psi}{2})\right).
\end{eqnarray}

Evidently, for a given setting of the path loss exponent $\beta$ and of $\gamma$ the above simple bound for the UL ergodic capacity depends only on the density of the network, i.e. the value of $\lambda$. Hence, if this bound could be extended in the general case of a complex network comprising both MCs and SCs, this could be proven a valuable tool for any  network operator and designer, toward meeting the QoS objectives based on the network densification. In this course, the UL ergodic capacity analysis is extended in the general case hereafter. 

\subsection{UL Ergodic Capacity Bounds - General Case}
\label{sec:DUDe}
Having defined the above ergodic capacity bounds for the cases of a single MC and of solely SCs, the analysis can now be extended to provide closed form simple expressions for the calculation of lower bounds for the cell ergodic capacity under a DUDe and a coupled association policy. In particular, the following general case analysis encompasses also, as particular cases, both the aforementioned approaches of Subsections \ref{sssec:MC} and \ref{sssec:SC}, as well as for the DUDe and the coupled association policy in complex HetNets.

In the coupled case the UE connects to the closest SC if the following holds for the distance $d$ to the SC:
\begin{eqnarray}
\label{eqn:Coupled_Crit}
d \leq \left(\frac{P_{SC}}{P_{MC}}\right)^\frac{1}{\beta} d_0,
\end{eqnarray}
where $P_{SC}$ is the transmit power of the SC and $P_{MC}$ is the transmit power of the MC. That is, the connection criterion of the UE is the level of the received power from each $AP$ \footnote{In the case of multi-antenna access points, the criterion in (\ref{eqn:Coupled_Crit}) should be rewritten as follows: $d \leq \left( \frac{M\rho(N)P_{SC}}{N\rho(M)P_{MC}} \right) ^{1/\beta} d_0$. This criterion is equivalent to connecting in the UL to the access point from which the highest rate is obtained in the DL.}.

In comparison, in the DUDe case the UE will connect to the closest $AP$ and not to the $AP$ from which it receives the highest power in the DL. Thus, the UE will connect to the closest SC instead of the MC if \footnote{In the case of multi-antenna access points, the criterion in (\ref{eqn:Crit2}) should be rewritten as follows: $d \leq \left( \frac{\rho(N)}{\rho(M)} \right) ^{1/\beta} d_0$. This criterion is equivalent to connecting in the UL to the access point so that the maximum UL rate is achieved.}
\begin{eqnarray}
\label{eqn:Crit2}
d \leq d_0. 
\end{eqnarray}

The previous two conditions can be unified under a single notation where the UE will connect to the closest SC if
\begin{eqnarray}
\label{eqn:alpha}
d \leq \alpha d_0,
\end{eqnarray}
where $\alpha = (\frac{P_{SC}}{P_{MC}})^\frac{1}{\beta}$ in the coupled case and $\alpha = 1$ in the DUDe case \footnote{Assuming that the antenna gains of the MC and the SCs are equal, whereas the $\alpha$ factor should be weighted accordingly if the antenna gains are different.}. Moreover, the above notation is general enough to account also for the cases of the Subsections \ref{sssec:MC} and \ref{sssec:SC}, since the case of a single MC corresponds to $\alpha = 0$ and the case of solely SCs corresponds to $\alpha = \infty$. Therefore, the selection criterion for the UL connectivity defined in (\ref{eqn:alpha}) takes all examined cases into account according to an \textit{a priori} defined value of $\alpha$. 

\subsubsection{Ergodic Capacity for the General Case}
In order to calculate the ergodic capacity for a generic network, encompassing both MC and SCs and supporting both DUDe and coupled transmission policies, a generic approach must be adopted taking into account the selection criterion described in (\ref{eqn:alpha}). In particular, the ergodic capacity can be calculated as the sum of the conditioned ergodic capacities in the case of the UE being connected to the MC and to the closest SC weighted by the probability of each of the two contingencies happening. Specifically, the average ergodic capacity is calculated as follows:
\begin{eqnarray}
\label{eqn:average}
\mathbf{E}[R] = \mathbf{E}_{h_0}[R|MC] P(MC) + \mathbf{E}_{h,d|SC}[R|SC] P(SC),
\end{eqnarray}
where $\mathbf{E}_{h_0}[R|MC]$ is the ergodic capacity conditioned to the fact that the reference user has connected to the MC for a given $d_0$, $P(MC)$ is the probability of the user to connect to the MC, $\mathbf{E}_{h,d|SC}[R|SC]$ is the ergodic capacity conditioned to the fact that the user has connected to the closest SC, and $P(SC)$ is the probability of the user to connect to the closest SC.

According to the selection criterion defined in (\ref{eqn:alpha}) the probability $P(SC)$ is equal to the probability $P(d \leq \alpha d_0)$, which after employing (\ref{eqn:CDF}) can be calculated by
\begin{eqnarray}
\label{eqn:PSC}
P(SC) = P(d \leq \alpha d_0) = ~~~~~~~~~~~~~~~~~~~~~~~~~~~~ \nonumber \\
\int_{0}^{\alpha d_0} 2 \pi x \lambda \exp(-\lambda \pi x^2) dx = 1 - \exp(-\lambda \pi \alpha^2 d_0^2),
\end{eqnarray}
and $P(MC)$ is calculated by
\begin{eqnarray}
\label{eqn:PMC}
P(MC) =  1 - P(SC) = \exp(-\lambda \pi \alpha^2 d_0^2).
\end{eqnarray}
Furthermore, in case $d > \alpha d_0$, a lower bound can be defined by (\ref{eqn:Bound}) and (\ref{eqn:Rayleigh}) for $\mathbf{E}_{h_0}[R|MC] $ as
\begin{eqnarray}
\label{eqn:EMC}
\mathbf{E}_{h_0}[R|MC] \geq \log(1 + d_0^{- \beta} \gamma \rho).
\end{eqnarray}

However, in order to compute lower bounds for $\mathbf{E}_{h,d|SC}[R|SC]$ a different approach than the one followed in (\ref{eqn:PDF}) needs to be considered, since the distance to the closest SC $d$ is conditioned by the fact that $d \leq \alpha d_0$ (i.e. it is conditioned by the fact that the UE has decided to connect to the SC). Therefore, the PDF defined in (\ref{eqn:CDF}) needs to be revised accordingly and the following truncated version of the PDF needs to be employed for the conditioned random variable $d|SC$ \cite{Papoulis}:
\begin{eqnarray}
\label{eqn:dSC}
f_{d|SC}(d|SC)= \left\{\begin{array}{l}0, ~~~~~~~~~~~~~~~~~~~~~~~ d < 0 \\
                               \frac{1}{k}~2 \pi d \lambda \exp(- \lambda \pi d^2), 0 \leq d < \alpha d_0 \\
				0, ~~~~~~~~~~~~~~~~~~~~~~~ \alpha d_0 \leq d \end{array}\right.
\end{eqnarray}
where $k$ is a constant selected appropriately so that the area of $f_{d|SC}(d|SC)$ is equal to $1$. That is, 
\begin{eqnarray}
\label{eqn:k}
k = \int_{0}^{\alpha d_0} 2 \pi x \lambda \exp(-\lambda \pi x^2) dx = P(SC).
\end{eqnarray}
Hence, similarly to (\ref{eqn:PDF}) the expected value $\mathbf{E}_{d|SC}[\log(d)]$ for the new PDF defined in (\ref{eqn:dSC}) can be calculated by
\begin{eqnarray}
\label{eqn:ESC}
\mathbf{E}_{d|SC}[\log(d)] = ~~~~~~~~~~~~~~~~~~~~~~~~~~~~~~~~~~  \nonumber \\
\frac{ \int_{0}^{\alpha d_0} \log(d) 2 \pi d \lambda \exp(-\lambda \pi d^2) dd}{P(SC)} = ~~~~~~~~~~~~ \nonumber \\ 
\frac{2 \int_{0}^{~\alpha d_0 \sqrt{\lambda \pi}} x \log(x) \exp(-x^2)dx}{P(SC)} - \log(\sqrt{\pi \lambda}).
\end{eqnarray}

Thus, after combining (\ref{eqn:average}), (\ref{eqn:PSC}), (\ref{eqn:PMC}), (\ref{eqn:EMC}), and (\ref{eqn:ESC}) the bound for the ergodic capacity in the general case is given in (\ref{eqn:General}) shown at the top of the last page. It is evident, based on the closed form expressions obtained from the preceding analysis, that the ergodic capacity in the UL depends only on the values of $\lambda$ and the distance $d_0$ from the MC access point, for a given UL association policy given by the decision factor $\alpha$.

\begin{cor}
For $\alpha d_0 \sqrt{\lambda \pi} \geq 4$, (\ref{eqn:General}) can be approximated by (\ref{eqn:General2}), which is given at the top of the last page. Thus, if the above criterion is met a simpler bound for the ergodic capacity in the UL can be employed. This approximation arises from the behavior of the integral: $\int_{0}^{~\alpha d_0 \sqrt{\lambda \pi}} x \log(x) \exp(-x^2)dx$, the value of which is approximately constant for any upper limit greater than 4. This can be further verified by the visual representation of the function that is being integrated depicted in Fig. (\ref{fig:integral}) of the Appendix.
\end{cor}

The simplicity of the derived analytical bounds as well as their dependency solely upon the the values of $\lambda$ and $d_0$ is of paramount importance for the network operator. In particular, these bounds provide complete information regarding the QoS and the densification of the network, enabling the network operator to adjust the network to the emerging traffic requirements. However, in order for these bounds to be of actual merit, and to provide an accurate picture of the network performance to the operator, they have to be tight. In the direction of corroborating how tight the obtained bounds are, the performance of a network comprising MC as well as SCs is simulated for all different settings defined above. Subsequently, these simulation results are compared  in the next Section against the analytical results obtained from the introduced bounds, verifying the tight relationship of both results.

\section{Simulations}
\label{sec:simulations}
In order to demonstrate the tight performance of the devised analytical bounds, a HetNet has been simulated encompassing a MC, SCs positioned according to a homogeneous PPP of spatial density $\lambda$ and a reference user. The SCs are located also beyond the coverage area of the MC. In addition, the simulations have been repeated for all four of the considered scenaria, i.e. for only a MC, for only SCs, for a HetNet under DUDe (i.e. $\alpha = 1$), and for a HetNet under a coupled association with $\alpha = 0.3$.

\begin{figure}
  \begin{center}
   \centerline{
    \includegraphics[height=7cm]{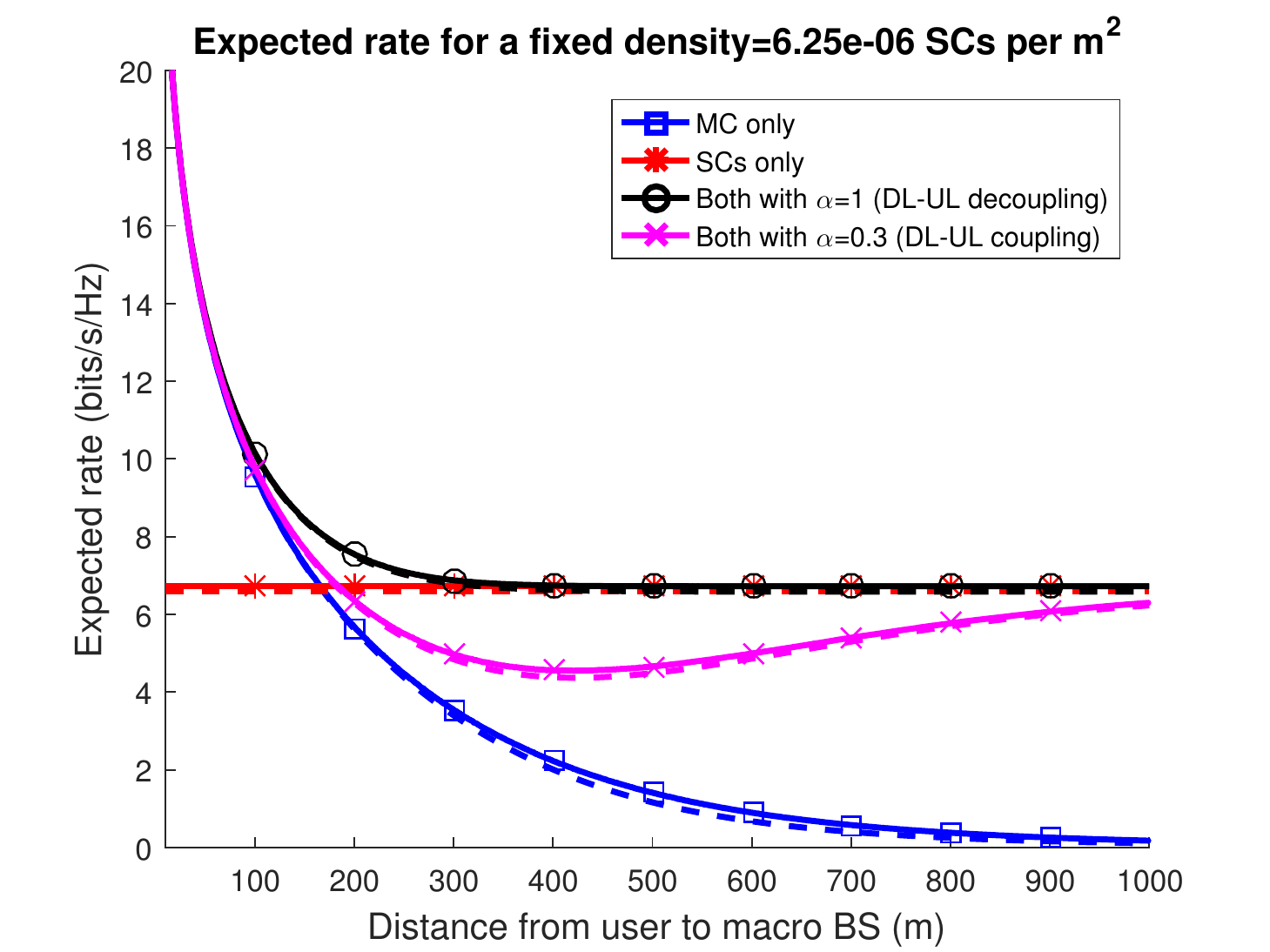}}
    \caption{Expected UL rate vs distance to the MC access point.}
    \label{fig:FixLam}
  \end{center}
\end{figure}

\begin{figure}
  \begin{center}
   \centerline{
    \includegraphics[height=7cm]{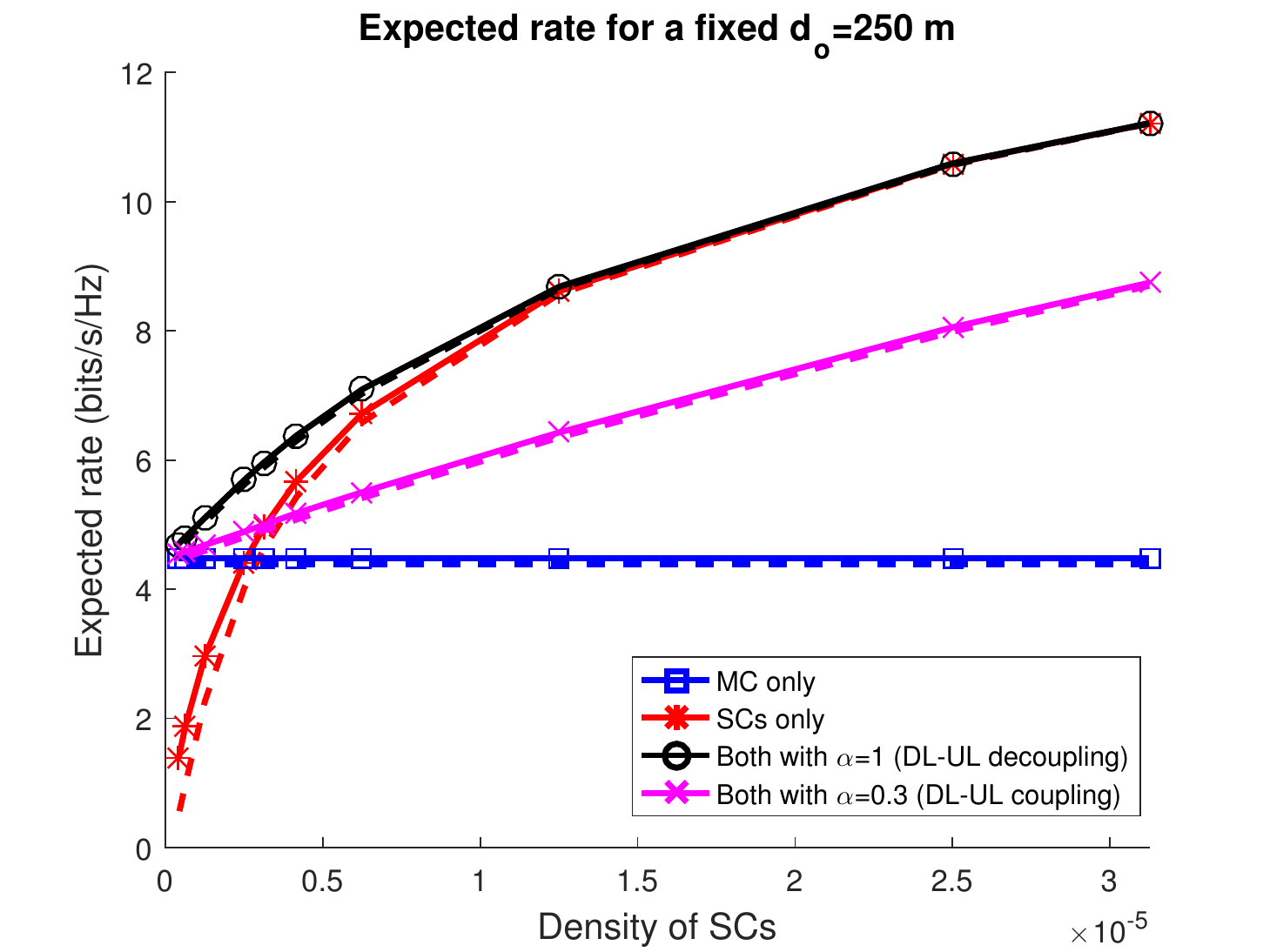}}
    \caption{Expected UL rate vs density of SCs.}
    \label{fig:FixD}
  \end{center}
\end{figure}

The basic parameters required for the link budget are tabulated in Table \ref{tab:Link_Bud}. According to these values the factor $\alpha = (0.01)^{0.25} = 0.3$ for coupled UL-DL association corresponds to a 20dB difference between the MC power and SC power, whereas $\alpha = 1$ corresponds to DUDe association policy and for both cases $\gamma$ is calculated based on the tabulated UE power. In the simulations a density $\lambda = 6.25~10^{-06} SC/m^2$ is defined \textit{a priori} and the expected rate has been compared against the distance $d_0$ from the MC access point with the performance of the simulated network and the analytical bounds being depicted in Fig. (\ref{fig:FixLam}). Subsequently, the simulations have been repeated for an \textit{a priori} defined distance $d_0 = 250m$ and the expected rate has been compared against the network density $\lambda$ in Fig. (\ref{fig:FixD}). Thus, the performance of the reference user is analyzed for a given distance from the single MC access point which is considered in the simulations. However, the analysis holds and can be extended in future work to quantify the performance of the reference user over the average of the distance $d_0$ from the MC access point. 

\begin{table}[h]
\caption{Link Budget Parameters}
\label{tab:Link_Bud}
\begin{center}
\begin{tabular}{|c|c|}
  \hline
  {\bf Parameter} & {\bf Value} \\ \hline
 \hline
  UE Transmit Power $P_{UE}$ &  33 dBm \\  \hline
  SC Transmit Power $P_{SC}$ &  33 dBm \\  \hline
  MC Transmit Power $P_{MC}$ &  53 dBm \\  \hline
  Bandwidth & 10 MHz \\  \hline
  Noise Power Spectral Density & -174 dBm/Hz \\  \hline
  Noise Power & -104 dBm \\  \hline
  Path Loss at Reference Distance $L_{ref} $ & 25.6 dB \\ 
  (Including Antenna Gains) &  \\  \hline
  Path Loss Exponent &  4 \\  \hline
  $\alpha$ (UL-DL coupling) &  $(0.01)^{0.25} = 0.3$ \\  \hline
  $\alpha$ (UL-DL decoupling) &  $1$ \\  \hline
\end{tabular}
\end{center}
\end{table}
The tight relationship between the obtained analytical bounds and the simulated results is manifested in both figures verifying the reliability of the preceding analysis and its utility in network management and design. In addition, another pivotal conclusion drawn from the presented simulations is the validity of the approximation result presented in (\ref{eqn:General2}), since the performance of the decoupled network converges to that of the network employing only SCs as the values of $\lambda$ and $d_0$ increase. A property arising from the fact that the performance of both networks is quantified by the integrated function of Fig. (\ref{fig:integral}), and therefore, the performance of both networks converges as the criterion defined in (\ref{eqn:General2}) is met. Hence, the simple bound presented in (\ref{eqn:General2}) can in fact be of actual merit to the network operator.

\begin{table*}[t] 
\begin{center}
\hrulefill
\begin{eqnarray}
\label{eqn:General}
\mathbf{E}[R] \geq \log(1 + \gamma d_0^{- \beta} \rho) \exp(-\lambda \pi \alpha^2 d_0^2) ~~~~~~~~~~~~~~~~~~~~~~~~~~~~~~~~~~~~~~~~~~~~~~~~~~~~~~~~~~~~~~~~~~~~~~~~~~~~~~~~~~~~~~~~~~~~~~~~~~~~~~~~~~~~~~~~~~~~~~~~ \nonumber \\
+ \log\left(1 + \gamma (\lambda \pi)^{\beta/2} \rho \exp \left( -\frac{2\beta}{1-\exp(-\lambda \pi \alpha^2 d_0^2)} \int_{0}^{~\alpha d_0 \sqrt{\lambda \pi}} x \log(x) \exp(-x^2)dx \right) \right)(1-\exp(-\lambda \pi \alpha^2 d_0^2)). ~~~~~~~~~~~~~~~~~~~~~~~~~~~~~~
\end{eqnarray}
\hrulefill
\begin{eqnarray}
\label{eqn:General2}
\mathbf{E}[R] \geq \log(1 + \gamma d_0^{- \beta} \rho) \exp(-\lambda \pi \alpha^2 d_0^2) + \log\left(1 + \gamma (\lambda \pi)^{\beta/2} \rho \exp \left( \frac{\beta \psi /2}{1-\exp(-\lambda \pi \alpha^2 d_0^2)} \right) \right)(1-\exp(-\lambda \pi \alpha^2 d_0^2)), ~~~~\alpha d_0 \sqrt{\lambda \pi} \geq 4
\end{eqnarray}
\hrulefill
\end{center}
\label{fig:Equations}
\end{table*}

\section{Conclusion}
\label{sec:conclusion}
In the present paper simple analytical expressions were provided for quantifying the performance of dense wireless networks under a DUDe association policy. Moreover, the obtained expressions provided a practical guide associating the network performance to the degree of densification of the network. The latter can facilitate the design and management of efficient cellular networks, where the QoS objectives can be guaranteed \textit{a priori} based on the spatial density of the network. Thus, this may provide an important leeway to the network operator to capitalize on the advantages of cellular dense networks.

Furthermore, the present paper sets the general framework for the extensions of the analysis in future work, to take the blockage effects on urban cellular networks into account, toward providing an interference free architectural paradigm for 5G ultra dense networks.

\appendix 
\renewcommand\thefigure{\thesection.\arabic{figure}} 
\section {Appendix}
\subsection*{~~~~~~~~~~~~~~~~~The function "$x \log(x) \exp(-x^2)$"}
\setcounter{figure}{0}    
\begin{figure}[!h]
  \begin{center}
    \includegraphics[height=7cm]{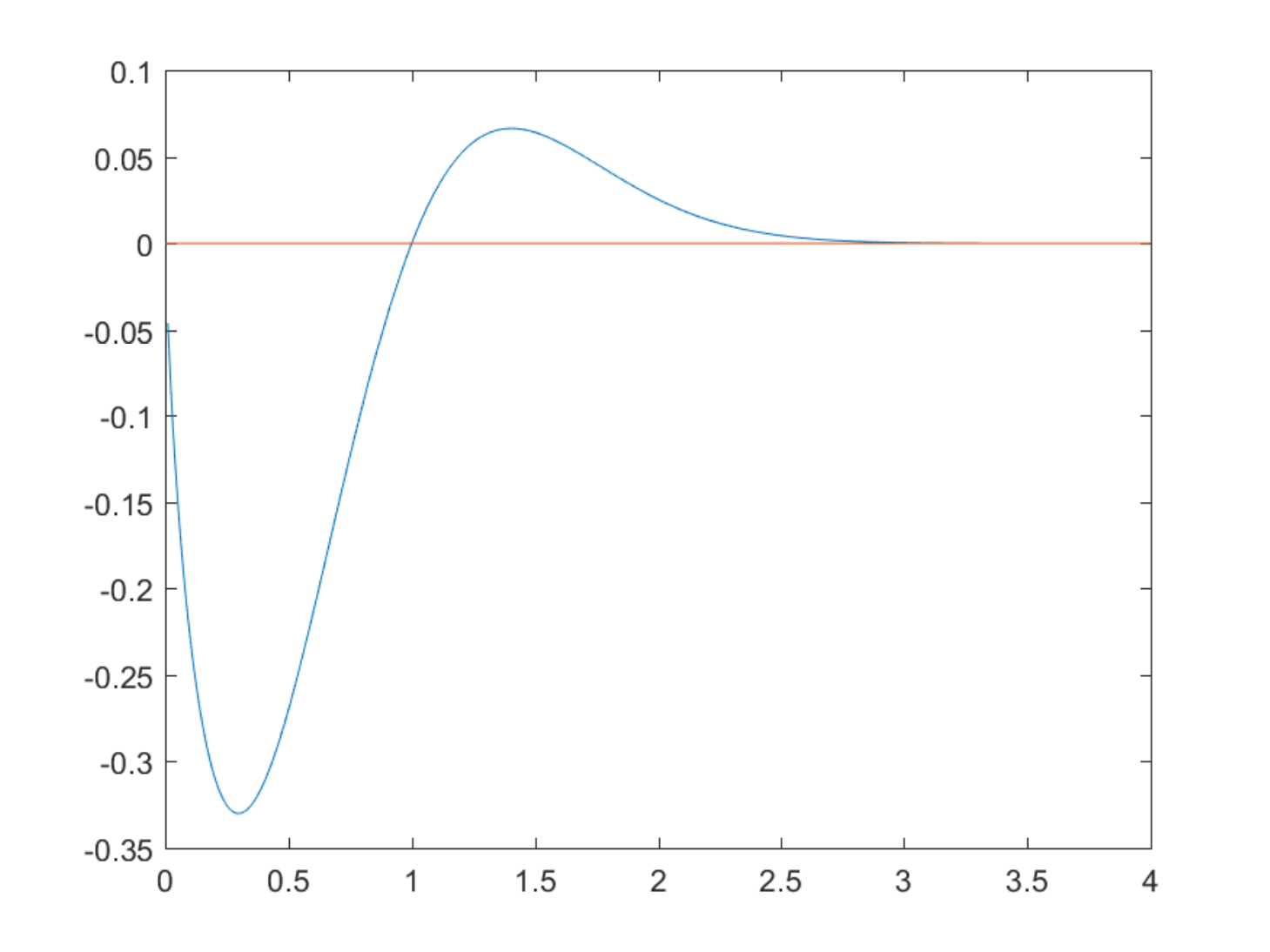}
    \caption{Visual representation of function $x \log(x) \exp(-x^2)$.}
    \label{fig:integral}
  \end{center}
\end{figure}

\section * {Acknowledgment}
\addcontentsline{toc}{section}{Acknowledgment}
The research leading to this paper has been carried out within the framework of project ETN-5Gwireless (this project has received funding from the European Union's Horizon 2020 research and innovation programme under the Marie Sk\l{}odowska-Curie grant agreement No. 641985) and project DISNET (funding from the Spanish Ministerio de Econom\'{i}a y Competitividad, project code TEC2013-41315-R). Additional funding has been received from the Catalan Government (AGAUR) through the grant 2014 SGR 60.

\balance 
\bibliographystyle{IEEEtran}
\bibliography{IEEEabrv,Camad.bib}

% \section*{Biographies}
% \begin{IEEEbiography}[{\includegraphics[width=1in,height=1.25in,clip,keepaspectratio]{img/\firstAuthorBioImage}}]{\firstAuthor} is blabla \end{IEEEbiography}
% \begin{IEEEbiography}[{\includegraphics[width=1in,height=1.25in,clip,keepaspectratio]{img/\secondAuthorBioImage}}]{\secondAuthor} is blabla \end{IEEEbiography}
% \begin{IEEEbiography}[{\includegraphics[width=1in,height=1.25in,clip,keepaspectratio]{img/\thirdAuthorBioImage}}]{\thirdAuthor} is blabla \end{IEEEbiography}
% \begin{IEEEbiography}[{\includegraphics[width=1in,height=1.25in,clip,keepaspectratio]{img/\fourthAuthorBioImage}}]{\fourthAuthor} is blabla \end{IEEEbiography}
% \begin{IEEEbiography}[{\includegraphics[width=1in,height=1.25in,clip,keepaspectratio]{img/\fifthAuthorBioImage}}]{\fifthAuthor} is blabla \end{IEEEbiography}

\end{document}